# A Comparative Analysis of High-Level vs. Low-Level Simulations for Dynamic MAC Protocols in Wireless Sensor Networks

Shama Siddiqui
*DHA Suffa University*
Karachi, Pakistan
South East Technological University
Waterford, Ireland
email: shamasid@hotmail.com

Anwar Ahmed Khan
*Millennium Institute of Technology & Entrepreneurship*
Karachi, Pakistan
South East Technological University
Waterford, Ireland
email: anwar.ahmed@mite.edu.pk

Indrakshi Dey
*Walton Institute for Information and Communication Systems Science*
South East Technological University
Waterford, Ireland
email: indrakshi.dey@waltoninstitute.ie

*Abstract*—Simulation studies are conducted at different levels of details for assessing the performance of Media Access Control (MAC) protocols in Wireless Sensor Networks (WSN). In the present-day scenario where hundreds of MAC protocols have been proposed, it is important to assess the quality of performance evaluation being conducted for each of the proposed protocols. It therefore becomes crucial to compare the results of high-level theoretical simulations with the detailed implementation results before any network protocol could be deployed for a real-world scenario. In this work, we present a comparison of high-level theoretical and detailed implementation results for Adaptive and Dynamic Polling-MAC (ADP-MAC). MATLAB has been used for conducting initial theoretical simulations and TinyOS has been used to develop the detailed implementation of protocol for Mica2 platform. Performance evaluation of ADP-MAC using the two levels of simulation has been conducted based on energy and delay. In the high-level implementation, energy consumption was found to be decreasing whereas delay was found to be increasing for increasing channel polling intervals. On the other hand, when detailed implementation was developed, it was observed that both energy consumption and delay revealed an increasing trend with the increasing polling intervals. Therefore, it has been shown that the trends for high- and low-level simulations for ADP-MAC are significantly different, due to the lack of realistic assumptions in the higher-level study.

## I. Introduction

Emerging applications of Internet of Things (IoT), such as smart city, wearable technology, home and industry automation and vehicular connectivity require state-of-the-art MAC layer protocols, which could facilitate optimal performance. Hundreds of MAC protocols for general as well as specific applications have been developed for the low power devices (Wireless Sensor Nodes & Networks) in IoT over the past two decades [1]. Due to the rapid emergence of industrial automation technologies and the role of IoTs, it has become crucial to identify the best suited MAC layer protocols [2]. Out of various performance parameters, energy and latency remain the hottest for managing the industrial communication networks.

The authors use various levels and tools for conducting performance assessment of their proposed MAC solutions. Often, authors present comparison of performance trends in MAC protocols using simulation, and testbed is rarely used due to the complications [3]. However, the match between the results of theoretical simulations and testbed experiments/low level simulations may not be possible to reach in every situation, particularly, when the details of WSN node or traffic characteristics are not possible to model completely [4]. Similarly, when the testbed experiments are conducted, the possibility of device malfunction and other real-world scenarios are better incorporated [5]. Despite high probability of missing out details in the high-level simulations, authors continue to present evaluation of their network protocols using the theoretical simulations [6], which creates hassles in transformation of the proposed protocols into real-world implementation. Therefore, there is a need to conduct comparison of the theoretical simulations results with those obtained after detailed testbed level implementation of the proposed protocols.

In the above context of studying the comparison of theoretical (high-level) simulations and testbed experiments (low-level), we present a comparison of results obtained for ADP-MAC. ADP-MAC has been developed with the motivation to timely serve the wireless sensor nodes through adjusting the polling interval distributions [7]. In comparison to the previous MAC protocols, which were designed based on altering the polling intervals [8], the novel contribution of ADP-MAC's design was to switch the 'polling interval distribution' instead of 'polling intervals'. Since no previous MAC was developed using polling interval distributions, the authors first tested the high-level idea of using interval distributions and simulations were performed at an abstract level using MATLAB [9]. Based on the satisfactory performance of initial results, the detailed implementation of ADP-MAC was developed for Mica2 platform, and the test-bed level simulation was conducted using Avrora. This paper describes the simulation results obtained through MATLAB & Avrora and offers a comparison of the two.

The primary motivation for this study is to address the limitations of relying solely on high-level simulations for evaluating MAC protocols, which often fail to reflect real-world conditions and challenges. High-level simulations provide initial insights but cannot account for practical issues like hardware constraints, packet collisions, and device malfunctions that affect real-world performance. By comparing the results of MATLAB simulations with testbed implementations on Mica2 motes, we aim to demonstrate the value of detailed, low-level evaluations in bridging this gap and ensuring the robustness of protocols like ADP-MAC for practical deployment.

Rest of this paper has been organized as follows: Section II presents a brief overview of the relevant work; Section III details the experimental set-up used for both simulation levels; Section IV presents the results and evaluation; finally, Section V concludes the paper.

## II. Related Work

Various adaptive MAC protocols have been developed for Wireless Sensor Networks in order to cater to the needs of recent applications of IoT [10]. One of the most efficient technique deployed in the adaptive MAC schemes is the dynamic duty-cycling where the nodes change their wake up and sleep duration based on the traffic arrival patterns [11]. Although these schemes have been quite successful, the protocols deploying dynamic channel polling (listening) have shown to even conserve more energy and adapt much better to the traffic requirements [12]. Hence, dynamic channel polling schemes are often required for the applications where energy conservation is of crucial importance, such as wireless body area networks [13].

An adaptive dynamic duty cycle mechanism for energy-efficient medium access control (ADE-MAC) has been proposed in [14] for Wireless Multimedia Sensor Networks (WMSNs). ADE-MAC employs an innovative asynchronous duty-cycle approach to manage the sleep patterns of sensor nodes, dynamically adjusting them based on the incoming traffic rate and queuing delays at each node; each node independently schedules its sleep patterns, requiring only the sender node to wake up the receiver nodes through the use of preamble packets. Although ADE-MAC significantly reduces synchronization overhead, it introduces increased delay due to the waiting time required for nodes to wake up. A variable duty cycle MAC (DC-MAC) [15] has taken synchronization approach by proposing a new method that only closely located nodes follow the same duty cycle, while the far-off nodes may follow a different. DC-MAC also has challenges due to synchronization overhead and high delay when need to send data to far-off nodes.

We proposed a dynamic channel polling scheme ADP-MAC in previous work [7]. The protocol was based on the idea that instead of altering the channel polling intervals as was the previous practice, polling interval distributions should be altered based on the analysis of the arrival distributions of traffic. For this purpose, we used statistical coefficient of variation (Cv) to identify the incoming arrival patterns. It was proposed that in case Cv is found to be higher than a certain threshold (0.8), exponential polling should be conducted, and deterministic polling should be used otherwise.

ADP-MAC is a duty-cycled asynchronous MAC in which the nodes begin their operation by sending preamble strobes, waiting for an early Ack, transmitting data packets and receiving final acknowledgement. In order to verify the impact of dynamic channel polling, we conducted performance evaluation for Constant Bit Rate (CBR) and Poisson arrivals and studied different types of polling: deterministic (channel polled at regular, pre-defined intervals), exponential (channel polled at exponentially distributed intervals) and dynamic (channel polling interval distributions switched between deterministic and exponential based on the traffic arrival patterns). Performance evaluation of ADP-MAC was conducted in two phases. Initially, MATLAB implementation was developed only to test the impact of varying polling interval distributions, and later a low-level TinyOS implementation was developed catering to the details such as collisions, preamble transmissions, retransmissions, etc.

A high-level study was conducted for the proposed dynamic channel polling mechanism in [9]. It was found that energy and delay performance both improve when exponential polling intervals are used for either CBR or Poisson arrivals. The rationale behind this finding was that when the channel is polled using exponential intervals, there is a higher probability of receiving aggregated packets. As a result, a single block acknowledgement could be used to send notification to the sender, which reduces the energy consumption. On the other hand, delay was found to be reduced for deterministic polling because packets could be received earlier due to having more regular polls.

In contrast, when the experiments were conducted for Mica2 using TinyOS implementation and Avrora simulator, it was found that energy consumption and delay both increase with increasing polling intervals. Also, the impact of arrival distribution was significantly visible over polling interval distribution: for CBR arrivals, it was found that deterministic polling serves best both in terms of delay and energy; for exponential arrivals, exponential polling served best and for bursty arrivals, dynamic polling was found to be the best choice. This finding was obtained because when the types of arrival and polling distributions match, there is a lesser delay between packet transmission and channel polling instant.

III. EXPERIMENTAL SET-UP

The study was conducted in two phases: High-level implementation using MATLAB and low-level implementation using Mica2 platform and Avrora simulator. The experimental set-up details for each platform have been explained below:

TABLE 1: MATLAB SIMULATION PARAMETERS

| Simulation Duration | 5000 secs |
|---|---|
| Mean inter-arrival duration | 5 secs |
| Mean polling interval | 1-10 sec |
| Size of Data Packet | 50Byte payload + 11Byte overhead |
| Size of Acknowledgement (ACK) Packet | 10B |
| Size of Preamble | 2B |
| Maximum no. of Concatenated in a Super packet | 5 |
| Energy consumed in Data transmission | 0.5 mJ/Byte |
| Energy consumed in Single Data packet transmission | 30.5 mJ |
| Energy consumed in ACK transmission | 5 mJ |
| Energy consumed in channel polling | 1 mJ |

*A. MATLAB Implementation Details*

At this stage, to quantify the impact of deterministic and exponential polling interval distributions, a high-level algorithm was written in MATLAB, instead of developing a full MAC protocol. In the preliminary experiments, arrays were generated to represent the packet arrivals and channel polling intervals for a single hop network. Assumptions were made for the values of packet sizes, energy consumption and delay involved for each transmission/reception activity, as presented in Table 1. (taken from [9]; most of these assumptions were made based on the base protocol's (Synchronized Channel Polling-MAC (SCP-MAC)) implementation [16].

Experiments were conducted for different combinations of exponential & CBR Arrivals and polling interval distributions. The mechanisms of packet concatenation and block acknowledgement were also included in the simulations; the packets which were received at a single poll were combined into a Super packet (packet concatenation) and a single acknowledgment packet was used to acknowledge the transmission of concatenated packets (block acknowledgements).

### B. Mica2 Implementation Details

ADP-MAC has been implemented in TinyOS [17] over the Mica2 motes [18]. TinyOS is an open-source operating system designed for embedded sensor networks. The Mica2 motes used in this research features AVR ATmega 128L chip – a microcontroller produced by Atmel. Instead of running the code on testbed physically, Avrora emulator [19] has been used, which has the convenience and flexibility of quickly setting up the network of different topologies and varying the number of nodes. Avrora is a cycle accurate instruction level sensor network simulator; the scalability of this simulator is up to 10,000 nodes, and it is 20 times faster than its contemporaries that offer a similar level of accuracy [20]. The experimental configuration parameters have been shown in Table 2. Each experiment was run 4 times and the results have been presented with a confidence interval of 95%.

TABLE 2: PARAMETERS USED FOR ADP-MAC'S DETAILED IMPLEMENTATION OVER MICA2

| Simulation Parameters | Value for ADP-MAC |
|---|---|
| Common Parameters | |
| Bit rate | 18.78 kbps |
| Arrival Patterns | CBR/Poisson |
| Polling Interval Distributions | Deterministic/Exponenti-al/Dynamic |
| Total Nodes | 10 |
| Message Generation Interval | 50 Sec |
| Number of packets transferred | 20 packets generated by each node |
| Distance between the Nodes | 1 m between each source and sink |
| Duration of Each Cycle $T_{cycle}$ | 10 sec |
| Threshold value of Cv | 0.8 |
| Size of Super Packet | Up to 5 data packets |

## IV. RESULTS AND EVALUATION

The results for energy consumption comparing the MATLAB implementation and Mica2 simulations have been presented in Figure 1, whereas Figure 2 shows the delay performance.

In the results obtained by high level implementation of ADP-MAC in MATLAB, it was observed that energy consumption decreases with the increase in polling intervals as shown in Figure. 1, where 1-a shows the performance of CBR arrivals, and 1-b presents the results for Poisson arrivals. On the other hand, the delay was seen to increase with increasing polling intervals as illustrated by Figure. 2; here Figure 2-a shows the delay trends for CBR, and 2(b) presents for Poisson arrivals; hence, a trade-off was found between the trends of energy consumption and delay, both for CBR and Poisson arrivals.

When the experiments for studying energy and delay performance of ADP-MAC were repeated based on the detailed low-level implementation of ADP-MAC, the trends obtained for energy consumptions have been shown in Figure. 1-c & 2-c. Here, it is to be noted that only trends can be compared but not the actual values; this is because in addition to lack of modeling real situation in the MATLAB implementation, there were also clear differences in the assumptions made for conducting MATLAB experiments and simulation parameters of detailed implementation of ADP-MAC

In contrast to the trade-off identified in the initial high-level simulations, there is no such trend seen in the detailed implementation results for ADP-MAC. Figure 1-c & 2-c illustrate that both the energy consumption and delay increase with the increasing polling intervals. Thus, the comparison for trends of energy consumption reveals an apparent contradiction between the initial high-level theoretical and detailed implementation results. However, for delay, both the initial as well as detailed implementation results reveal the same (increasing) trend with the increase in polling intervals.

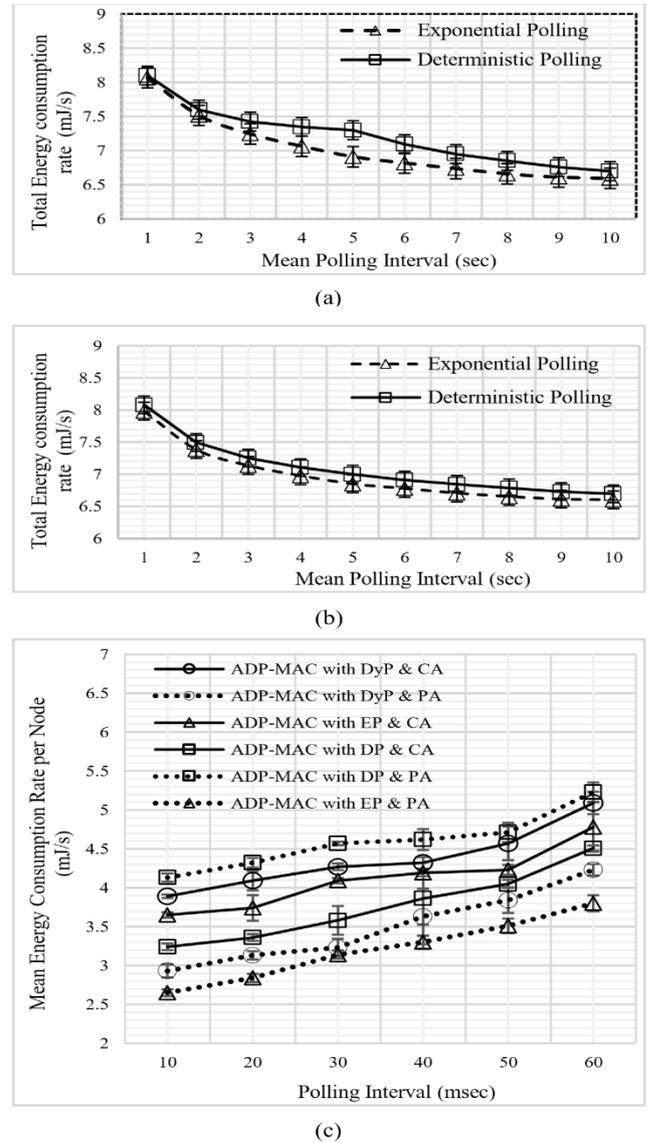

Figure 1: Total Energy Consumptoon (a) High Level Results for CBR Arrivals (b) High Level Results for Poisson Arrivals (c) Results for Detailed Implementation

The second apparent difference between the MATLAB & TinyOS implementation results is that it was predicted by MATLAB results that Exponential Polling should result in better energy performance for both the arrival processes, i.e. CBR and Poisson. Similarly, the delay performance was expected to be better for the Deterministic Polling regardless of the arrival process. In contrast, the findings of detailed implementation have revealed that for each arrival distribution type, if the polling interval distribution type is similar, it results in better performance for both energy consumption as well as for delay; i.e. for CBR Arrivals, the Deterministic Polling results in better energy and delay performance, whereas for Poisson Arrivals, Exponential Polling has come out to be a better candidate.

The rationale behind the first difference between the expected and detailed implementation results is the fact that the high-level prediction results are based on several assumptions, which are not entirely valid in the detailed implementation. For example, in the MATLAB implementation, the energy consumption was calculated based on the assumptions about the level of energy consumed in polling activities and data & ACK transmissions. For both the deterministic and exponential polls, the mean number of polls were always shown to be the same with only a change in their distribution. The energy savings was depicted through the transmission of reduced bytes due to packets received as concatenated and block acknowledgements. However, there was no implementation of the preamble transmissions, collisions, Carrier Sense Multiple Access/Collision Avoidance (CSMA/CA) process and retransmissions; all these details have been taken into account in the actual implementation of the ADP-MAC, which have resulted differently due to the assumptions about realistic channel and network conditions.

Moreover, since the preamble transmission mechanism was not modeled in the theoretical results, the energy consumption showed decreasing trend as at the higher polling intervals, the number of polls reduced, and the block acknowledgements increased leading to the better energy performance. However, in the implementation, as the polling interval increases, the preamble transmissions increase; this increases the energy consumption as although the polling energy will reduce for this case but at the same time, the preamble transmission energy and collisions would increase. This happens because the channel remains occupied for longer intervals, as compared to the cases where the polling interval was set to a low value and the packet could be quickly transmitted.

The second difference about the trade-off shown between the energy and delay in the MATLAB results can also be justified. In the implementation, no such trade-off exists, and the energy and delay reveal similar increasing or decreasing trends for both the arrival patterns. This happens because for the cases where the energy consumption would be higher due to the preamble transmissions, excessive polling, collisions and retransmissions, the delay will also be higher and vice versa. This insight was hard to predict in the high-level results due to the lack of consideration of the details of preamble transmissions and the possible collisions and retransmissions.

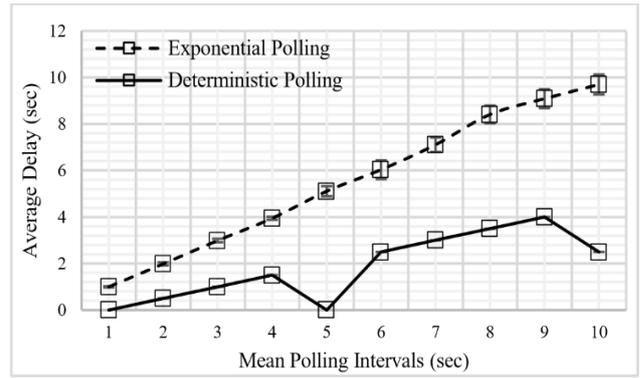

(a)

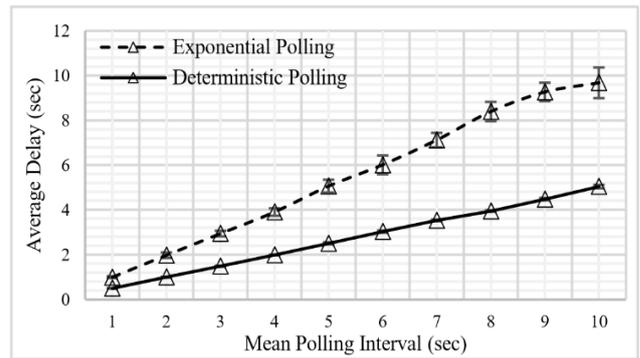

(b)

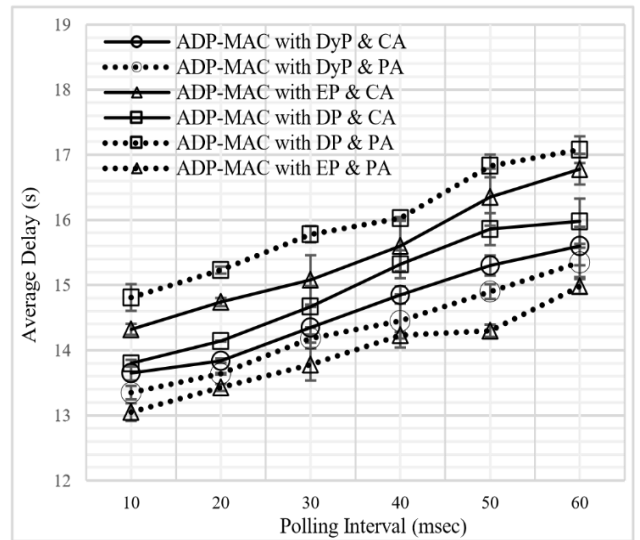

(c)

Figure 2: Average per Hop Delay (a) High Level Results for CBR Arrivals (b) High Level Results for Poisson Arrivals (c) Results for Detailed Implementation

On the other hand, the results for delay in both the high level as well as detailed implementation remain same; the increasing trend for delay has been observed with the increase in polling intervals as in both the cases, the packet would be transmitted when the poll would take place.

In light of the above results, it has been observed that the trends for energy consumption and delay in ADP-MAC differ between theoretical simulations and testbed experiments (low-level simulation). This finding highlights the need for caution when relying solely on simulation studies for real-world application implementation. It is possible that discrepancies arise due to factors such as potential bugs in the simulation program, limitations in the analytical model (e.g., not accounting for all energy consumption factors), or differences in implementation quality that may not have been captured in the simulations. Testbed-level studies, on the other hand, incorporate real-world scenarios, including hardware malfunctions and network dynamics, which theoretical simulations cannot fully replicate. Thus, researchers are encouraged to conduct both simulation and testbed experiments to validate their network protocols and ensure robust real-world performance.

## V. LIMITATIONS AND CONSTRAINTS

Based on the experimental setup, results, and evaluations, several limitations, boundaries, and constraints were identified. First, the high-level MATLAB simulations rely on several assumptions, such as fixed energy consumption values and simplified channel conditions, which may not reflect real-world scenarios, leading to discrepancies in the results when compared with detailed testbed implementations. Additionally, the lack of modeling preamble transmissions, collisions, and retransmissions in the MATLAB implementation led to unexpected energy consumption trends when applied to the Mica2 motes using TinyOS and Avrora. Furthermore, the use of an emulator (Avrora) instead of a physical testbed introduces limitations in capturing real-world sensor network dynamics, such as hardware malfunctions and environmental factors, which can influence the overall system performance. While the study used a reasonable range of assumptions and experimental setups, scalability to larger networks and diverse environmental conditions remains a boundary that should be explored in future work. Finally, the trade-offs between energy consumption and delay observed in the testbed implementation suggest that different network configurations may produce different results, highlighting the need for context-specific protocol adjustments.

## VI. CONCLUSION AND FUTURE WORK

This paper presented the results of performance evaluation conducted for an adaptive and dynamic MAC protocol (ADP-MAC). The authors initially conducted high-level simulation for the protocol using MATLAB, whereas the detailed implementation was developed at a later stage using Mica2 platform and Avrora simulator. Significant differences were observed in the trends of energy and delay when the two simulation results are compared. The rationale behind these differences in trends for the two simulation levels reveals that the high-level simulations may not reveal the true performance of the protocols as various predictions/assumptions do not remain valid for the full implementation. Also, it is not possible to model all the attributes of physical layer in the high-level simulations. For example, while conducting the initial study, the concepts of channel access delays and packet collisions could not be modeled. Based on these findings, it is suggested that the proposed protocols should be implemented to the node level in order to reflect close to real performance.

In future, we plan to implement ADP-MAC in large-scale, real-world testbeds to evaluate its performance under diverse network conditions and traffic scenarios. Moreover, we also aim to explore enhancements that address challenges such as heterogenous traffic management, potentially through emerging machine learning-based optimization techniques.

ACKNOWLEDGMENT

This contribution is supported by HORIZON-MSCA-2022-SE-01-01 project COALESCE under Grant Number 10113073.